\documentstyle[aps,prd,eqsecnum,epsf,preprint]{revtex}
\begin{document}
\preprint{\baselineskip=18pt \vbox{\hbox{SU-4240-630(revised)}
\hbox{February,1997}}}
\title{Deconfinement Transition and Flux-String Models}
\author{Arshad Momen and Carl Rosenzweig}
\address{Department of Physics, Syracuse University,\\
 Syracuse, NY 13244-1130, U.S.A.}
\date{\today}
\maketitle
\begin{abstract}
Flux-string models can be used to study the deconfining phase transition. 
In this note, we study the models proposed by Patel. We also study the 
large $N_c$ limits of Patel's model. To discuss the validity 
of the mean field theory results, 
the one-loop Coleman-Weinberg effective potential 
is calculated for $N_c=3$. We argue that the quantum corrections
vanish at large $N_c$ when the energy of the so-called baryonic vertices  
scale with $N_c$.
\end{abstract}
\newcommand{\be}{\begin{equation}}
\newcommand{\ee}{\end{equation}}
\newcommand{\bea}{\begin{eqnarray}}
\newcommand{\eea}{\end{eqnarray}}
\newcommand{\bt}{\beta}
\newcommand{\tp}{\tilde{p}}
\newcommand{\tH}{\tilde{h}}
\newcommand{\tJ}{\tilde{J}}
\newcommand{\s}{\sigma}
\newcommand{\th}{\theta}
\newcommand{\ra}{\rightarrow}

\section { Introduction}
The structure of the  quantum chromodynamics (QCD) the vacuum  
has been a subject of intense investigations 
over the last two decades \cite{shuryak}, as qualitative
understanding of hadron 
spectroscopy depends on the vacuum configuration.
Since the QCD vacuum confines color degrees
of freedom, it is highly 
non-trivial compared to the vacuum for QED.
However, to address this issue in a sensible way from the 
fundamentals of QCD is a rather challenging task as 
the theory becomes strongly 
non-perturbative in the infrared region. 

It is customary to use 
effective Lagrangians in the low energy domain to describe
the hadronic dynamics using the global symmetries of the theory.
However, these models are inadequate for studying
the vacuum structure since the order
parameters - the mesonic fields,  are 
color singlets.

Therefore, one has to take a path somewhere in between, explicitly
keeping some of the colored 
degrees of freedom, like quarks, 
while truncating the other degrees of freedom, namely 
the gluons. 
These truncations are performed 
in various disguises -  via potentials,
bags, strings etc. One can also argue in favor of 
the roles played by topologically non-trivial classical solutions  like 
instantons and subsequently perform a semiclassical
analysis \cite{callan} around these solutions. 

An economical picture for confinement is provided by 
the Meissner effect in  type-II superconductors. It is 
argued that the behavior of QCD vacuum is  dual to the 
type-II superconductors \cite{dual,rev}.
In this picture, the chromoelectric field is screened in 
the ``dual superconductor'' just like magnetic field in ordinary 
superconductors. Therefore the chromoelectric field between a quark  
and an anti-quark gets squeezed in a tube-shaped region like the 
Abrikosov flux-tubes in type-II superconductors.
This stringy picture for mesons dates back to the 
Nielsen-Olesen model \cite{string}, where the existence of 
a scalar field is crucial. But, apparently in QCD no such 
scalars exist. According to 't Hooft 
 the role of the scalars can be mocked up in
QCD via a suitable gauge choice and 
the so-called ``Abelian Projection'' \cite{abelian}.
Evidences  for  this type of  string models come from the Regge behavior
\cite{regge} and 
 the hadronization processes in  
high energy hadronic reactions where it gives 
a rather consistent picture
\cite{lund}. 

It is possible to incorporate baryons in this picture as well 
\cite{vene}. To do so, one has to include special configurations
called the 
``Y-vertices'', which allows $N_c$ colored quarks to bind and hence form 
baryons. These vertices are exclusive 
to the models  with 
$N_c > 2$ and the flux-strings can form  branches in these cases. 
Evidence for appearance of such objects in the 
magnetic condensation picture has also been argued by using 
Makeenko-Migdal equations \cite{olesen}.
It has also been 
argued that these vertices are gluonic in nature and  trace 
the baryon number \cite{kharzeev}. One is allowed to associate 
some energy with these vertices which is not associated with the
flux-tube itself.

On the other hand, at high temperatures
it has been argued from general
principles \cite{polyakov,susskind} that a  deconfining 
phase in QCD exists. An important 
issue is then the nature of the  phase transition. The order of
phase transition can be found by using universality arguments and 
looking at appropriate spin models which belong to the 
same universality class as QCD \cite{SY} for 
various color groups. In these spin models, 
the phase transition proceeds via the formation of 
domains at lower temperatures by cooling the system from a high 
temperature phase where one is guaranteed a disordered state
which can be understood as the deconfined phase.

One can use the flux-string models to study the same problem, 
though here one adopts a bottom-up approach - i.e. we start 
at lower temperature in the confining phase and then raise 
the temperature of the system. This forces the flux-string  
to  fluctuate and at high temperatures the string will elongate 
and make a criss-crossed mesh. At the deconfining point this net 
will permeate the whole space. 
This way of addressing the deconfinement transition in a flux-string
model was studied  
by Patel \cite{patel1,patel2}. 

In the original work of Patel \cite{patel2}, a flux-string model was
proposed on the lattice where the strings lived on the links and the 
quarks and the baryonic vertices were defined on the lattice sites
subject to the Gauss law for the conservation of $U(1)$ color charges
modulo $N_c$.
The model is simple enough to be studied analytically yet it retains much 
of the relevant physics in a model. 
Patel used mean field theory to evaluate the 
free energy of the 
system. In this treatment, the system showed  phase-transition of second 
order for $N_c =2 $ and of first order when $N_c=3$ provided 
one takes into account the baryonic ``Y-vertices''.
 This result qualitatively
agrees with the result found by
looking at the center of the gauge group \cite{SY}.
 The difference between $N_c=2 $ and 
$N_c= 3$ arises from the presence of the so-called 
``Y-vertices'' in the string picture.  
In fact, this is consistent with the model of 
Patel. 

In fact, the gluonic vertices resolve the following puzzle.
In absence of these vertices and quarks, the phase transition 
predicted by Patel's model is of second order, independent of the 
value of $N_c$. This originates  
from the fact that the 3D pure XY model has a second order 
phase transition \cite{3dxy}, although some subtleties with 
this result exist in the literature \cite{jahnke} ( see also 
\cite{stone}). The universality 
arguments of  \cite{SY} on the other hand  show that 
 the QCD phase transition for $N_c=3$   
is of first order, even in the case of pure gauge fields. 
Therefore, to resolve this apparent conflict, it is natural to include 
these vertices in these models and require them to be gluonic in nature.

At first sight, Patel's results might seem a little unsettling 
as they rely on a mean field treatment as one is 
aware of the role of fluctuations in lower dimensions. However, as we will
show below, this result is correct provided the results are interpreted 
in terms of large $N_c$. This is 
the issue we would like to address in this paper. 

We take the following route. 
The model that Patel gets on the lattice is the 
non-linear U(1) $\sigma$ model in three dimensions with some 
specific type of symmetry breaking. This model has a
physical cutoff built into it, namely the thickness of the 
string. We evaluate the one-loop
effective potential for this model in the continuum. We find that there
are contributions which are non-analytic in the coupling constants.
Being non-analytic they can't be absorbed by any finite choice
of counter-terms and hence must be included in the initial 
theory. Therefore, we include the non-analytic terms 
in the free-energy expression and hence perform an ``improved''
mean-field theory calculation. 

The paper is arranged as below. In section 2, we discuss the behavior
of these baryonic vertices with increasing $N_c$. Next in section 3,
we discuss the flux-string model for the simplest case 
of $N_c=2$ where all hadrons can be represented simply as unbranched  
flux tubes and correspondingly the problem can be mapped into a 
non-backtracking random walk on a cubic lattice. We also review 
the model with
$N_c \geq 3$  where the flux-tubes can branch and this warrants a different 
treatment from the $N_c=2$ case. In section 3, we give details of the
mean field calculation done by Patel. 	In section 4,we discuss the naive 
continuum limit for the model of section 3 and discuss the phase structure
using the continuum model. In section 5, we calculate the one-loop effective 
potential using the Coleman-Weinberg results and find the non-analytic 
contribution to the tree level effective potential. The non-analytic piece 
is then included and the mean field free energy 
is computed for the modified system.

\section{Baryonic Vertices at large $N_c$}

\begin{figure}
\epsfxsize=4in
\epsfysize=4in
\epsfbox{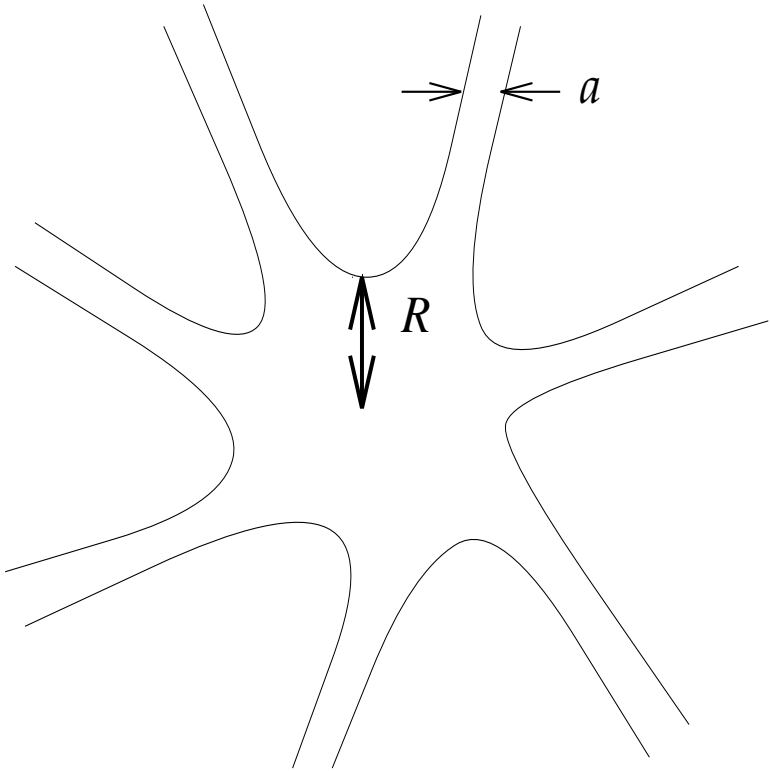} 
\caption{Baryonic vertices with finite volume}
\label{fig1}
\end{figure}

The string model can also serve as a model for large $N_c$ QCD. In both 
models baryons are qualitatively differnt from mesons. In string theory 
this difference is coded in the $N$-string vertex. Witten \cite{baryon}
has argued that the mass associated with the vertex increase with $N$. 
We will incorporate this behavior in our picture. 
The flux tube generated in three color QCD has a   
finite cross sectional  area. This area $a^2$ is expected to shrink 
as $\frac{a_0^2}{N}$ in the limit as $N$ goes to infinity, where $a$ is the 
width of the string and $a_0$ is a constant. Since in the string 
model of a baryon there are $N$ strings converging on the vertex the total 
cross sectional area of all the strings is finite. Thus there is finite region
surrounding the vertex, of radius $R \sim a_0$, where the strings overlap
( see Fig.\ \ref{fig1}).

It is this  region of finite volume which provides the common background 
interaction for all the quarks ( with the strings) that is the essence of 
Witten's large $N$ picture of the baryons. According to 
Witten \cite{baryon} this background interaction is 
responsible for a contribution of O($N$) to the baryon mass. Since, this
junction of finite area is present for $N_c$ , we incorporate 
this contribution as the effective mass $v$ of the $N$-string junction and 
thus have $v \ra N$ as $ N \ra \infty$. 

It is quite interesting that the above picture for baryons can be merged nicely
with the Skyrmion picture for baryons. Since these flux tubes repel each 
other ( like their type-II superconductor counterpart), it is only  
natural that at large $N_c$, baryons would look like hedgehogs with 
quarks stuck at the ends of each ``needle''. We believe this picture 
is quite parallel to the case when one can interpolate between the 
bag or string picture for mesons \cite{VPCR}.

\section{ Flux Tube Models for Deconfinement}
\subsection{SU(2) flux-strings : Random Walk}
 
We review the flux-string picture of 
deconfining transition  by  describing the simplest case, 
when the color gauge group is $SU(2)$.
For $SU(2)$ gauge theory, flux tubes represent both
mesons (made of a quark and an anti-quark) and
baryons ( made of two quarks). On the lattice,  these flux strings
can be 
represented by locally non-backtracking random walks \cite{patel1}. 
So, they are 
different from self-avoiding random walks, as they can cross each other 
after several steps. 
The Hamiltonian for a flux tube of length $l$ is given by 
\be
H = \s a l
\label{a1}
\ee
and  the grand partition function is 
\be
Z(\bt)= \sum_{l} e^{-(\bt \s l a)}
\label{a2}
\ee
For non-back tracking motion on a $d$-dimensional hypercubic lattice, 
there are $(2 D -1)$ different directions  that the flux tube can move to 
and hence for a flux tube of length $l$ in lattice units one has 
$(2 D - 1)^l$
different configurations when the ends of the flux tube is allowed to be 
free. Hence the grand partition function is \footnote{
 one can include a  chemical potential for quarks with finite masses, though 
we will ignore it hereafter.}
\be
Z(\bt) = \sum (2 D -1)^l e^{- \bt \s l a} = \sum e^{-l [ \bt la - \ln (2d-1)]}.
\label{a3}
\ee
 The partition 
function diverges, which is a typical signature for phase transitions,
at the temperature  -
\be
k_B T_c = \frac{\s a }{\ln (2d -1)}.
\label{a4}
\ee
In fact, this is a second order phase transition. This model has recently
been discussed by Kiskis also \cite{kiskis}.

\subsection{Case for $N_c \geq 3$ }

The more general model proposed by Patel \cite{patel1,patel2}
can be described as follows. 
In this model one can include flux-string branchings which 
are present in models with  $N > 2$, and hence this is more general 
than the model presented above. 

As before, we work on a D-dimensional hypercubic lattice. The fluxtubes 
are described via the the variables $n_{i,\mu}$ living on the $\mu$-th link
on the lattice site $i$ which are valued in the set 
 $\{0,\pm 1\}$. Also one has quarks occupying the lattice sites 
described by the variable $p_i \in \{0, \pm 1\}$ and the baryonic vertices 
described by the variable $q_i \in \{0,\pm 1\}$. The Hamiltonian describing 
the system is given by
\be
H= \sum_i [\s a \sum_\mu n_{i,\mu}^2 + m p_i^2 + v q_i^2 ]
\label{b1}
\ee
where $v$ is the energy associated with each $N$-legged vertex representing
the baryons. However, as in the gauge theory, one also has the Gauss law 
constraint
\be
\sum_\mu ( n_{i,\mu}-n_{i-\mu,\mu}) -p_i + Nq_i \equiv \alpha_i= 0
\label{b2}
\ee
which we need to implement at each lattice site. Hence the partition function 
is
\be
Z(\s a, \bt) = \sum e^{-\bt H}\prod_i \delta_{\alpha_i,0}
\label{b3}
\ee
The partition function can be evaluated by rewriting the delta function 
as
\be
\delta_{\alpha_i,0}= \frac{1}{2\pi}\int^{2\pi}_0 d \th_i e^{i\alpha_i \th_i}
\label{b4}
\ee
This allows us to express our partition function as 
\be
Z = \prod_i \int \frac{d\th_i}{2\pi} \sum_{n_{i,\mu}}
e^{-(\bt \s a n_{i,\mu}^2 
+ in_{i,\mu}(\th_{i+\mu,\mu}-\th_{i,\mu}))}
\sum_{p_i}(e^{-(\bt m p_i^2 + i p_i 
\th_i)})\sum_{q_i}e^{-(\bt v q_i^2 + i Nq_i \th_i)}
\label{b5}
\ee
Carrying out the summation over the variables $n_{i,\mu}, p_i , q_i$ we get
\be
Z(\bt) =\prod_i \int \frac{d \th_i}{2\pi}
\prod_\mu ( 1+ 2e^{-\bt \s a}
\cos (\th_{i+\mu}-\th_i))
(1+2 e^{-\bt m}\cos \th_i)(1+2e^{-\bt v}\cos(N\th_i))
\label{b6}
\ee
However, we are interested in the limit  $\bt
 \rightarrow \infty$ and hence it is expedient to replace terms like 
$(1+2e^{-\bt m}\cos \th_i)$ by $e^{h \cos \th_i}$ where $h = 
2e^{-\bt m}$. The partition function is then given by
\be
Z(\bt)= \int \prod_i d\th_i e^{J \sum_{\mu,i}\cos (\th_{i+\mu}-\th_i)+
h\sum_i \cos \th_i + p \sum_i\cos (N\th_i)}
\label{b7}
\ee
where
\be
J= 2e^{-\bt \s a} \qquad h=2e^{-\bt m} \qquad p=2e^{-\bt v}
\label{b8}
\ee
When the baryon 
vertices are very massive, i.e. $p \rightarrow 0$, 
the expression (\ref{b7}) reduces to the partition function of the 
D dimensional  XY model with an external magnetic field.

The mean field free energy can be calculated ( as in Appendix 1)
and one gets \cite{patel2},
\be
\bt F \leq -[ \ln I_0 (z) + D J (\frac{I_1(z)}{I_0 (z)})^2 + (h-z) \frac{I_1(z)}
{I_1(z)} + p \frac{I_N(z)}{I_0(z)}]
\label{c10}
\ee

Let us now deal with the case $N_c=3$ in detail.
For small values of $z$ it is sufficient to study the critical
parameters  
(\ref{c10}) using a series expansion  in $z$ for the free energy
\be
\bt F_{m.f.} = -\frac{h z}{2} + \frac{z^2}{4} ( 1- DJ) - 
\frac{z^3}{48}(p -3h) - \frac{3 z^4}{64} ( 1-\frac{4D}{3}J) + O(z^5).
\label{c11}
\ee
Using this and assuming small values for $p$ and $h$ one can 
easily find the critical parameters \cite{patel2}.
When the quarks are infinitely massive ( i.e. $ h= 0$) one finds
\be 
1- DJ_0 = \frac{p^2}{36}
\label{c12}
\ee
where  $J_0$ is the critical parameter at the 
which the system shows phase transition.
When $p=0$ , one can see $J_0 = \frac{1}{D}$ and the system shows 
a {\it second 
order} phase transition.

However, when quarks have some finite mass,  $h \neq0$  and 
the critical parameters  are found to be 
\be
1- D J_0 = \frac{p^2}{24} \qquad \qquad  h_{cr}= \frac{p^3}{216}.
\label{c13} 
\ee
When $h > h_{cr}$, the system shows no phase transition. 

The above  mean field treatment shows that the model exhibits 
a  second order phase transition 
which is the case for $N_c = 2$ and this result seems in agreement with 
the results of $\cite{SY}$. However, as one knows mean field treatment 
is not reliable in $d > 4$, one should treat 
this result rather cautiously.

Before we attempt to discuss the stability of the phase structure against 
quantum fluctuations, it is interesting to discuss the phase transition in 
this type of model for arbitrary $N_c >3$.

\begin{figure}
\epsfxsize=5.5in
\epsfysize=6in
\epsfbox{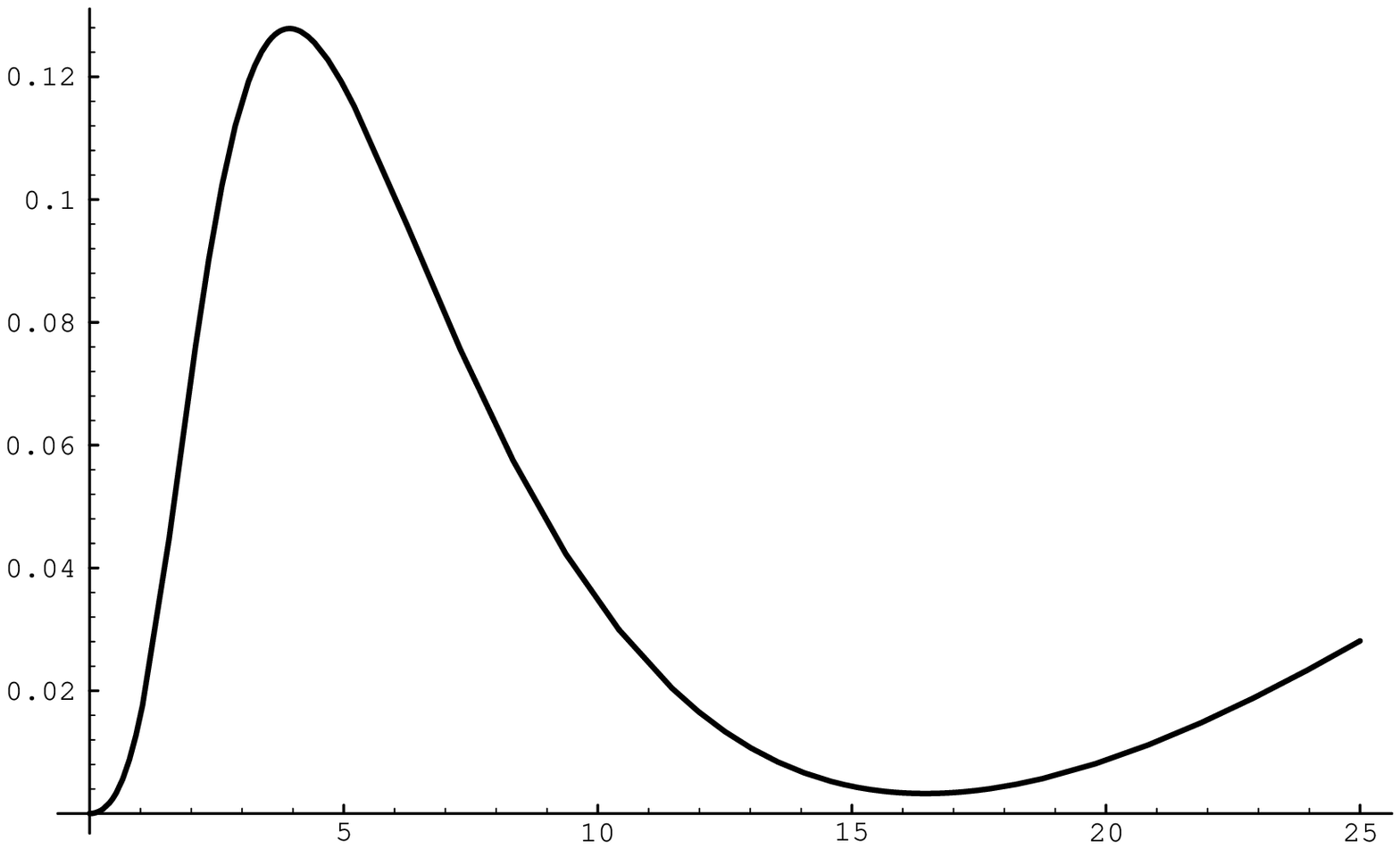}
\caption{Free Energy as a function of $z$ for $N_c=4,~p=1.48,~J=.32$}
\label{fig2}
\end{figure}

\begin{figure}
\epsfxsize=5.5in
\epsfysize=6in
\epsfbox{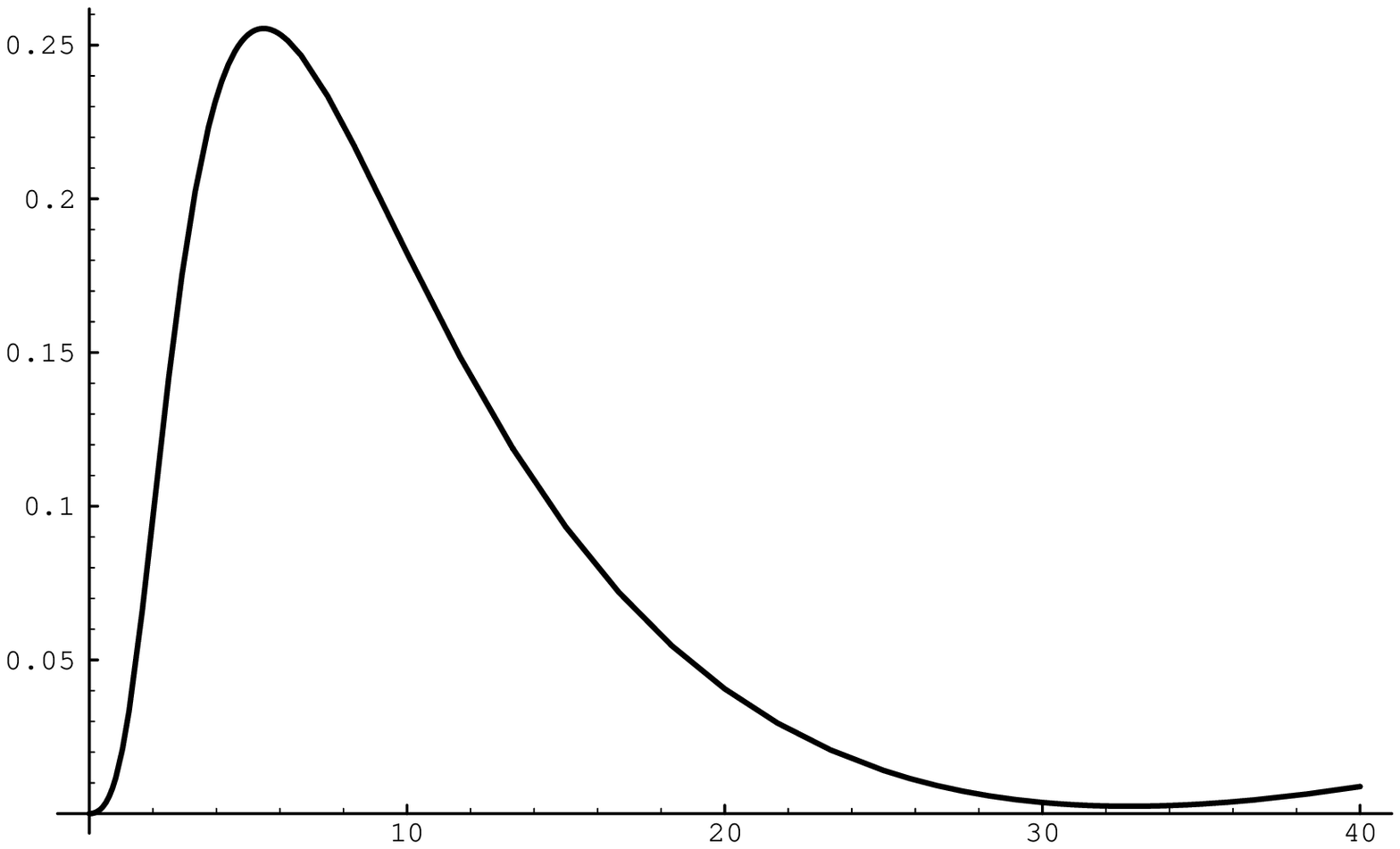} 
\caption{Free Energy as a function of $z$ for $N_c=5,~p=1.8,~J=.32$}
\label{fig3}
\end{figure}

\begin{figure}
\epsfxsize=5.5in
\epsfysize=6in
\epsfbox{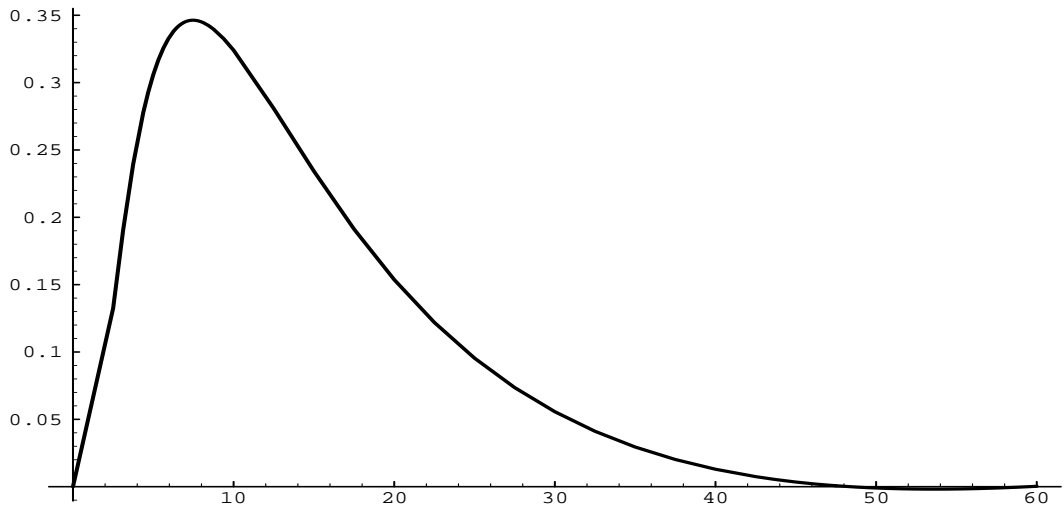}
\label{fig4} 
\caption{Free Energy as a function of $z$ for $N_c=6,~p=2,~J=1/3$}
\end{figure}

 In the second paper of Patel \cite{patel2} it was claimed that 
the order of the phase transition depends on the value of $p$. When $h=0$,
it was argued that for low values of $p$ the phase transition is of second 
order and at some high values of $p$ it is of first order. This can be 
checked explicitly by plotting the mean field free energy given by 
(\ref{c10})( see Fig.\ \ref{fig2},\ref{fig3},\ref{fig4}, showing 
the critical cases). 
Though this is, in principle, true, 
we don't consider the result reliable. 

The second vacua appears at large values of $z$ and for $p$ near its maximum 
 value of $2$. This critical value of $p$ seems to increase 
as $N_c$ increases ( see Figures). On the contrary large $N_c$ 
argument imply that $p$ actually decreases as $N_c$ increases.
Thus for large $N_c$, implying small $p$ we expect only a second order phase 
transition. Given a physically reasonable value of $p$, (for $N_c = 3$),
$~ p< 1$, Patel's model seems to imply a second order phase transition 
for $N_c \geq 4$, in
agreement with \cite{SY}. This result also supports the recent arguments of 
\cite{pisarski} that a second order phase transition for $N_c \geq 4$ would 
provide insight into the weakly first order QCD deconfinement transition
for $N_c=3$. In our context, $N_c=3$ is somewhat special.

\section{Reliability of  Mean field Results}

To check validity of the mean field results, we have to compare
 one-loop contribution to the free energy with the mean field result. 
However, as it stands the Hamiltonian is awkward for such a 
perturbative calculation, 
since the interaction terms are nonpolynomial.
The way to get around this is to notice that one can rewrite the Hamiltonian 
as the nonlinear $O(2) \sim U(1)$ chiral model. Defining $U= e^{i \th}$ 
\be
-\bt H
= \sum_i \frac{1}{2}\left[ \sum_\mu J (U_i U^{\dagger}_{i + \mu} + h.c. ) +
h ( U_i + U^{\dagger}_i ) + p ( U^N_i + U^{\dagger N}_i ) 
\right]
\label{d1}
\ee
Notice that the quark masses appears like the chiral symmetry breaking term 
as in the standard 
chiral Lagrangians but we also 
have the extra term due to the baryonic vertices .
We also have to impose the constraint 
\be
U U^\dagger = 1
\label{d2}
\ee  
At long wavelengths we can replace the first term in equation (\ref{d1})
by the derivative so that the Hamiltonian reads 
\be
-\bt H= \int d^3x \frac{1}{2}\left[ -{\tJ}  \nabla
U \cdot \nabla U^{\dagger}  +
\tH ( U + U^{\dagger}) + \tp ( U^N + U^{\dagger N} ) 
\right]
\label{d3}
\ee
where we have scaled the various coupling constants as follows.

We will be interpreting the above model as an effective theory with a
short distance cutoff $a_0$ which is  the thickness of the string. This 
cutoff will 
be present as the fluxtube description breaks down at distances shorter 
than the string thickness. In terms of $a_0$ one can see $\tp = \frac{p}{a_0^3}
$, $\tH = \frac{h}{a_0^3}$ and $\tJ = \frac{J}{a_0}$. We will restore this 
cutoff later.

Even when $\tp , \tH =0$ this model is nontrivial due to its 
nonlinear structure and  shows a first order phase transition 
which is 
understood as a vortex loop condensation \cite{vortexcondensation}.
This model also has an interesting
 large $N_c$ behavior as can be understood
 as follows.
The quantum contribution to effective action for this model involves 
 $(N-2)$th order 
loop and hence has a factor of $\hbar^{N-2}$. As $N_c \ra \infty$, 
this goes to zero and hence the theory is  expected 
to become classical.
 
Hereafter  we will restrict ourselves to the case {\underline{
$N=3$ only}}. 
Writing 
\be
U= (\phi + i \psi) ; \qquad \phi^2 + \psi^2 = 1
\label{d4}
\ee
we get after eliminating $\psi$
\be
 \bt H =\int d^3x \left[\frac{1}{2} \tJ \frac{1}{1-\phi^2}|\nabla
\phi|^2   -
(\tH -3\tp) \phi -  4 \tp \phi^3   
\right]
\label{d5}
\ee
 Note that in terms of variables $\th$ , $\phi = cos \th$.
The potential term
\be
V(\phi) \equiv - (\tH -3\tp) \phi -  4 \tp \phi^3   
\label{d6}
\ee
admits a local minimum only if $3\tp > \tH$, which is given by 
\be
\phi_0 = - \sqrt{\frac{ 3\tp - \tH}{12\tp}}\equiv - v
\label{d7}
\ee  
Lets expand the ``Hamiltonian'' (\ref{d5}) about this vacuum,
\be
\phi \equiv - v + \sigma
\label{d8}
\ee
where $\sigma$ is the fluctuation about the vacuum $v$.
Then one has 
\be
\bt H = \int d^3 x [ - 4 \tp v^3 + \frac{J}{2(1-v^2)} \nabla \sigma 
\cdot \nabla \sigma - 12 \tp v \sigma^2 - 4  \tp \sigma^3 ]
\label{d9}
\ee
Rescaling the field $\sigma = \sqrt{\frac{9 \tp + \tH}{12 \tp \tJ}} \chi$ 
one can rewrite the ``Hamiltonian'' for the 
fluctuation  (after dropping the constant piece) as  
\be
 \bt H'  = \int d^3x [ \frac{1}{2} | \nabla \chi |^2 + \frac{1}{2}
M^2 \chi^2 - \frac{1}{3!}g \chi^3 ]
\label{d10}
\ee
where 
\bea
M^2 \equiv \frac{9\tp+\tH}{\tJ} \sqrt{1-\frac{\tH}{3\tp}} \nonumber \\
g \equiv \frac{1}{\sqrt{3\tp}} (\frac{9\tp+\tH}{\tJ})^{\frac{3}{2}}
\label{d11}
\eea
Note that when $\tp = \frac{\tH}{3}$ the fluctuations become massless.

\section{Coleman-Weinberg Effective Potential}
Though the nature of phase transition can be found by looking at the classical
potential, the presence of fluctuations might change the structure in lower
dimensions , i.e. mean field calculations based on Gaussian fixed points are 
not reliable. To find the deviations, it is customary to look at the one-loop
effective potential, as calculated first by Coleman and Weinberg \cite{Cole}.
Given a classical potential $V(\phi)$ , the  effective potential up to one-loop
is given by 
\be
V_{one-loop}(\phi) = V(\phi) +\frac{1}{2}\int \frac{d^3k}{(2\pi)^3 } \ln ( 
G^{-1}(k) + \frac{\partial^2 V(\phi)}{\partial \phi^2})
\label{e1}
\ee
For our case , $V(\phi)$ is as in (\ref{d6}) and the inverse 
propagator $G^{-1}(k)$ is given by $  \frac{Jk^2}{1-\phi^2}$
so that one-loop contribution to the the effective potential reads,
\be
V_1(\phi) \equiv  \frac{1}{2}\int \frac{d^3k}{(2\pi)^3 } \ln (\frac{Jk^2}
{1-\phi^2} - 24 \tp \phi )
\label{e2}
\ee
The finite contribution from this integral can be evaluated by 
the dimensional regularization method, details of which can be found in the 
appendix. The answer is 
\be
-\frac{\hbar}{12\pi}[W(\phi)]^{\frac{3}{2}} 
\label{e3}
\ee
where
\be
W(\phi)\equiv -\frac{24 \tp }{\tJ}\phi ( 1 -\phi^2)
\label{e4}
\ee
Note that the one-loop contribution is real when $\phi$ is negative.
In terms of the original variables namely $\th$, this one-loop 
contribution reads
\be
-4\sqrt{6}\frac{\hbar}{\pi}(\frac{\tp}{\tJ})^{\frac{3}{2}}\cos^{\frac{3}{2}}
\th \sin^3 \th 
\label{e5}
\ee
This expression is non-analytic and hence can not be absorbed 
by a suitable counter. Hence, we follow the following 
strategy - namely, include this term in the outset and hence redo 
the mean field calculation. Basically, we are then working with an effective
action which includes the non-analytic term. 
Restoring the cutoff one sees that 
the relevant term to add in the lattice theory 
would be 
\be
-4\sqrt{6}\frac{\hbar}{\pi}(\frac{p}{Ja_0^2})^{\frac{3}{2}}\cos^{\frac{3}{2}}
\th \sin^3 \th. 
\label{e5a}
\ee
We would like comment on this quantum correction in the 
large $N_c$ limit. It is known that \cite{flux} in large $N_c$ limit 
the thickness of the strings goes to zero with $a_0 \sim \frac{1}{\sqrt{N_c}}$.
So, naively the contribution to this term increases with $N_c$. However, 
if $v \sim N_c$ then one sees that this correction goes to zero as
$N_c \ra \infty$. Hence, the theory becomes classical, which is consistent with
our naive expectations.

In terms of the lattice theory, the inclusion of this term 
adds the following term in (\ref{c6})
\be
-4\sqrt{6}\frac{\hbar}{\pi}(\frac{p}{Ja_0^2})^{\frac{3}{2}} \langle \cos^{\frac{3}{2}} \th \sin^3 \th \rangle 
\label{e6}
\ee
However, this contribution is non-analytic in  $z=cos \th$. Hence,
we have to take a branch-cut along the positive $z$-axis
and take the positive root above the cut and the negative root 
below the cut. Then, one can readily evaluate the above integral, details of 
which we will relegate to the appendix, as  
\be
-8\sqrt{6}\frac{\hbar}{\pi}(\frac{p}{Ja_0^2})^{\frac{3}{2}} \frac{\pi}{I_0(z)}[ \frac{5}{2} { _1F_1}(\frac{5}{2},\frac{7}{2};z)
 - \frac{9}{2}
{ _1F_1}(\frac{9}{2}, \frac{11}{2};z) ]
\label{e7}
\ee
As a series expansion in $z$, this reads 
\be
-\frac{32\sqrt{6}\hbar}{\pi^2}(\frac{p}{Ja_0^2})^{\frac{3}{2}} 
[ 0.022 + 0.013 z - 0.0013 z^2 - 0.0022 z^3 
- .00188 z^4 + O(z^5)]
\label{e8}
\ee
this term will now contribute to the  
mean field free energy given by (\ref{c11}). Defining $ \lambda \equiv 
\frac{32 \sqrt{6} \hbar}{\pi^2} (\frac{p}{Ja_0^2})^{3/2}$, the total free
energy is given by
\bea
\bt F = -0.022 \lambda - z( \frac{h}{2} + .013 \lambda) 
&+& z^2 ( \frac{1-3J}{4} 
+ .0013 \lambda) - z^3 ( \frac{p-3h}{48} -.0022 \lambda) \\ \nonumber
&-& z^4
( \frac{3(1-4J)}{64} - .00188 \lambda ) + \cdots.
\label{e9} 
\eea
The behavior is this free energy function is similar to the mean field 
result and one can show that the phase transition do not change.

\section*{conclusions}

In this paper, we have attempted to incorporate quantum fluctuations
in a flux-tube model by looking at a continuum limit of the model. The 
effect of the quantum fluctuations are interesting though they do not change 
the nature of the deconfining phase transition. However, to show this one 
has to use the fact that the ``Y-vertices'' has to scale with $N_c$. Only 
then is the picture consistent with large $N_c$. The large $N_c$ limit of 
Patel's model implies that for $N_c \geq 4$, the phase transition is of second
order.

\section*{Acknowledgements}
We would like thank the members of the high energy
theory group at Syracuse University for discussions.
A.M. wishes to thank especially 
P. Benetatos, M. Harada , G. Jungman and M.C. Marchetti. This work was 
supported by the US Department of Energy contract number DG-FG02-85ER40231.

\appendix

\section{Mean Field Calculation}

Here we reproduce the Free energy expression for the model using the 
mean field calculation. Remember that our model is defined in three dimensions
and hence mean field calculations are not reliable. 
 Then the partition 
function can be written as 
\be
Z_0 =  \int \prod_i d{\bf S}(\th)_i e^{J \sum_{\mu,i} {\bf S}_i \cdot 
{\bf S}_{i+\mu}+
h\sum_i {\bf B}\cdot {\bf S}_i }
\label{c1}
\ee
To evaluate the partition function, we now resort to mean-field where the 
spins are assumed to produce a mean magnetic field $z$ so that the partition 
function is 
\bea
Z(\bt)= \int \prod_i \frac{d\th_i}{2\pi}
 e^{z\cos \th_i}  e^{J \sum_{\mu,i}\cos (\th_{i+\mu}-\th_i)+
(h -z)\sum_i \cos \th_i + p \sum_i\cos (N\th_i)} \nonumber \\
= \left[ I_0 (z)\right]^N \langle e^{J \cos ( \th_{i+\mu} - \th_i) 
+ h cos \th_i + p \cos (N\th_i) }\rangle_{M.F.}
\label{c2}
\eea
where $I_N (z)$ is the Bessel function with imaginary arguments of $N$th 
order and $\langle \cdots \rangle_{M.F.}$ is the expectation value evaluated 
using the mean field weighing factor,i.e.
\be
\langle A \rangle_{M.F.} \equiv \frac{\prod_i\int \frac{d \th_i}{2\pi}
 A(\th) e^{z \cos \th_i}}
{\prod_i \int \frac{d \th_i}{2\pi}  e^{z \cos \th_i}}
\label{c3}
\ee
 Now we use the fact for any 
convex function one has the following inequality 
\be
\langle e^A \rangle \geq e^{\langle A \rangle }
\label{c4}
\ee
Hence the partition function is bounded below by the following quantity
\be
Z(\bt) \geq [ I_0(z)]^N e^{\langle J \cos (\th_{i + \mu} - \th_i ) 
+ (h-z) \cos \th_i + s \cos (N\th_i) \rangle }
\label{c5}
\ee
Recall that the free energy expression is given by $ F =\lim_{N \rightarrow 
\infty} -\frac{1}{\bt N} \ln Z$ and hence 
\be
\bt F \leq \bt F_{M.F.} = - \left[\ln [ I_0 (z)]+  \langle J \sum_\mu
\cos ( \th_{i +\mu} 
- \th_i ) + (h-z) \cos \th_i + s \cos (N \th_i) \rangle\right] 
\label{c6}
\ee  
Now notice that $\langle \sin \th \rangle= 0 $ and thus 
\be
\bt F \leq -\left [\ln [ 2 \pi I_0 (z)]+ 3 \langle \cos \th_i \rangle ^2 + 
(h-z) \langle \cos \th_i \rangle + s \langle \cos N \th_i \rangle \right]
\label{c7}
\ee
The various averages can be carried out easily by noting
\bea
< \cos N\th > &=& 
\frac{\int^{2\pi}_{0}d\th e^{\alpha \cos \th} \cos{N\th}}{\int^{2\pi}_{0}
e^{\alpha\cos \th}} \nonumber \\
&=& \frac{\int^{2\pi}_{0}d\th \sum_{n=-\infty}^\infty I_n ( \alpha) \cos n\th
\cos N\th}{\int^{2\pi}_{0} d\th \sum^{\infty}_{n=- \infty} I_n(\alpha) \cos 
n\th}
\label{c8}
\eea
which gives us the result ,
\be
< \cos{N \th} > = \frac{I_N (\alpha)}{I_0(\alpha)}.
\label{c9}
\ee
Using this is (\ref{c8}), we get the expression (\ref{c10}).

\section{The One-loop contribution to the free energy}

In this appendix we give the details of our various computations. 
From (\ref{e2}), one can see that thew one-loop contribution can be 
written as 
\be
V_1(\phi) = \frac{\hbar}{2} 
\int \frac{d^3k }{(2\pi)^3} [ \ln ( k^2 - 24\frac{\tp}{\tJ} \phi ( 1- \phi^2))
- \ln (\frac{ (1-\phi^2)}{\tJ})].
\label{ap1}
\ee
The second term in the integrand can be dropped in dimensional regularization.
Defining $W(\phi) \equiv - 24 \frac{\tp}{\tJ} \phi (1 - \phi^2)$,
 one can find 
\bea
\frac{\partial V_1(\phi)}{\partial \phi } &=& \frac{\hbar W'(\phi)}{2} 
\int \frac{d^3k }{(2\pi)^3} \frac{1}{k^2 + W(\phi)}= \frac{\hbar W'(\phi)}
{16 \pi^{3/2}} \Gamma (-\frac{1}{2}) \sqrt{ W(\phi)} \\ \nonumber
&=& - \frac{\hbar}{8 \pi} \sqrt{W(\phi)} W'(\phi)
\label{ap2}
\eea
with $W'(\phi) = \frac{\partial W}{ \partial \phi}$. Integrating both sides
 and dropping an irrelevant constant, we get 
\be
V_1 ( \phi ) = - \frac{\hbar}{12 \pi} [ W(\phi)]^{3/2} = - \frac{ 4 \sqrt{6}
\hbar}{\pi} ( \frac{\tp}{\tJ})^{3/2} \{ \phi ( 1 - \phi)\}^{3/2}.
\label{ap3}
\ee 
Putting in $ \phi = \cos \th $ one gets the answer (\ref{e3}).

Next we  calculate the mean-field contribution due to $V_1 ( \phi)$. Now,
\bea
< \cos^{3/2} \th \sin^3 \th > &=& \frac{ \int^{2 \pi}_0 d\th \, 
e^{z \cos \th } cos^{3/2} \th \sin^3 \th }{ \int^{2 \pi}_0 e^{z \cos \th}} \\
\nonumber
&=& \frac{1}{2 \pi I_0(z)} \sum^{\infty}_{m=0} \frac{z^m}{m!}( \int^{2 \pi}_0
 d \th \cos^{3/2 +m} \th \sin \th  - \int^{2 \pi}_0 d\th 
 \cos^{7/2 + m } \th 
\sin \th ).
\label{ap4}
\eea
 The integrals above can be performed by noting that $\sqrt{\cos \th}$ is not 
an analytic function and we have to introduce a branch-cut as stated 
previously. We choose $\sqrt{\cos \th} = 1$ for $\th = 0 $ and $\sqrt{
\cos \th}= -1$ for $\th = 2\pi$. Then we can readily evaluate
\bea
<\cos^{3/2} \th \sin^3 \th > &=& \frac{1}{ \pi I_0(z) }
\sum_{m=0}^{\infty} \frac{z^m}{m!} \{ \frac{1}{ m + 5/2} - \frac{1}{ m + 9/2}
\} \\ \nonumber 
&=& \frac{2}{\pi I_0 (z) } [ \frac{1}{5} { _1F_1} (5/2, 7/2, z) - \frac{1}{9}
{ _1F_1} (9/2, 11/2, z) ]
\label{ap5}
\eea
which leads us to (\ref{e7}).
\newpage
\newpage
\newpage
\newpage

\end{document}